\documentclass{aa}  
\usepackage{graphicx}
\usepackage{txfonts}
\begin{document}

    \title{A study of the prompt and afterglow emission \\
       of the Short GRB 061201\thanks{The results reported in this 
paper are partially based on observations carried out at ESO telescopes 
under program 078.D-0809}}


   \author{G. Stratta \inst{1}, P. D'Avanzo \inst{2,3}, S. Piranomonte \inst{4}, S. Cutini \inst{1,5}, B. Preger \inst{1}, M. Perri \inst{1}, M.L. Conciatore\inst{1,6}, S. Covino \inst{3}, L. Stella \inst{4}, D. Guetta \inst{4}, F.E. Marshall \inst{8}, S. T. Holland \inst{8,9}, M. Stamatikos \inst{8},  C. Guidorzi \inst{3,10}, V. Mangano \inst{11}, L. A. Antonelli \inst{1,4},  D. Burrows \inst{7},  S. Campana \inst{3}, M. Capalbi \inst{1}, G. Chincarini \inst{3,10}, G. Cusumano \inst{11}, V. D'Elia \inst{4}, P.A. Evans \inst{12}, F. Fiore \inst{4},  D. Fugazza \inst{3}, P. Giommi \inst{1}, J.P. Osborne \inst{12}, V. La Parola \inst{10,11}, T. Mineo \inst{11}, A. Moretti \inst{3}, K.L. Page \inst{12}, P. Romano \inst{3,10}, G. Tagliaferri \inst{3}
          }

   \offprints{G. Stratta}

   \institute{ASI Science Data Center, via Galileo Galilei, 00044, Frascati, Italy\thanks{INAF personnel resident at ASDC} \\
   \and
	    Universit\`a degli Studi dell'Insubria, Dipartimento di Fisica e Matematica, 
		via Valleggio 11, I-22100 Como, Italy\\
   \and      
             INAF--Osservatorio Astronomico di Brera, Via E.\ Bianchi 46, I-23807 Merate (LC), Italy\\
   \and
		INAF--Osservatorio Astronomico di Roma, 
              via Frascati 33, 00040, Monte Porzio Catone, Italy\\
   \and
		Universit\`a di Perugia, Dipartimento di fisica,
 	      Viale A. Pascoli, I-06123 Perugia, Italy\\
   \and
		Universit\`a degli studi di Roma ''La Sapienza'', 
              Piazzale Aldo Moro 2, I-00185, Roma, Italy\\
   \and
		Department of Astronomy and Astrophysics, Pennsylvania State University, 
                525 Davey Lab, University Park, PA 16802, USA\\
  \and
		Astrophysics Science Division, Code 660.1, NASA/GSFC, 
		Greenbelt, MD 20770, USA\\
 \and
	   Universities Space Research Association, 10227, Wincopin Circle, 
	   Suite 221, Columbia, MS 21044, USA\\
   \and
             Universit\`a{} degli Studi di Milano, Bicocca, 
             Piazza delle Scienze 3, I-20126, Milano, Italy\\
  \and
	    INAF Istituto di Astrofisica Spaziale e Fisica Cosmica, Sezione di Palermo, 
	    via Ugo la Malfa 153, 90146, Palermo, Italy \\
  \and
            Department of Physics \& Astronomy, University of Leicester,
		Leicester LE1 7RH, UK\\
		}


 
  \abstract
   {
Our knowledge of the intrinsic properties of short duration Gamma-Ray Bursts has relied, so far, only upon a few cases for which the estimate of the distance and an extended, multiwavelength monitoring of the afterglow have been obtained. 
   }
   {
We carried out multiwavelength observations of the short GRB 061201 aimed at estimating its distance and studying its properties.  
   }
   {We performed a spectral and timing analysis of the prompt and afterglow emission and discuss the results in the context of the standard fireball model. }
{

A clear temporal break was observed in the X-ray light curve about 40 minutes after the burst trigger. We find that the spectral and timing behaviour of the X-ray afterglow is consistent with a jet origin of the observed break, although the optical data can not definitively confirm this and other scenarios are possible. No underlying host galaxy down to $R\sim26$ mag was found after fading of the optical afterglow. Thus, no secure redshift could be measured for this burst. The nearest galaxy is at z=0.111 and shows evidence of star formation activity. We discuss the association of GRB 061201 with this galaxy and with the ACO S 995 galaxy cluster, from which the source is at an angular distance of  17$''$ and 8.5$'$, respectively.  We also test the association with a possible undetected, positionally consistent galaxy at $z\sim1$. 
In all these cases, in the jet interpretation, we find a jet opening angle of $1-2$ degrees. 

}
   {}


\authorrunning{Stratta et al.}

\titlerunning{A study of the prompt and afterglow emission of the 
       Short GRB 061201}

   \maketitle
%

\section{Introduction}

Short duration Gamma-ray Bursts (GRBs) are historically defined as those GRBs with burst duration less than two seconds and hard spectra (Kouveliotou et al. \cite{kouve1993}). As the sample of short GRBs increases, in order to take into account all the spectral and temporal properties of the prompt emission, alternative empirical definitions of short GRBs have been introduced (e.g. Norris \& Bonnell \cite{norris2006}; Zhang et al \cite{zhang2007}). 

The expected progenitor for short GRBs is a merging binary system of compact objects. However, 
our present knowledge of the intrinsic properties of short bursts mainly relies upon a few cases for which the distance could be derived and an extended multiwavelength monitoring of the afterglow was carried out. 
From these it appears that short bursts are less energetic and less collimated than long GRBs (e.g.  Fox et al. \cite{fox2005}, Burrows et al. \cite{burrows2006}, Soderberg et al. \cite{soderberg2006}). Intrinsic spectral parameters such as the peak energy $E_p$ of the $E F(E)$ spectrum of the burst and the equivalent isotropic energy $E_{\rm iso}$, do not match the  $E_{\rm p}-E_{\rm iso}$ correlation for long GRBs (e.g. Amati \cite{amati2007}).

The short bursts for which unambiguous hosts have been found are at redshifts between $z=0.10$ and $z=0.55$. The optical afterglows of seven short bursts and their likely host galaxies have been recently observed with telescopes such as ESO/VLT, Gemini, Magellan and HST, but no firm estimation of the redshift could be obtained. The apparent faintness of the likely host galaxies  indicates that they are at redshift $z\ge0.7$ and consequently that about $25\%$ of the short bursts revealed in the Swift/HETE-2 era may reside at larger distances than found so far (Berger et al. \cite{berger2007}).

So far, for all short GRBs, host galaxy candidates with various statistical significance have been found. In particular, every well localized short GRB ($<5''$) has a candidate host galaxy consistent with negligible offset.
In this work we present a multiwavelength study of a short GRB detected by Swift (Gehrels et al. \cite{gehrels2004}), GRB 061201, for which the  optical afterglow was clearly detected, but no underlying galaxy was found. 
We investigate the possibility that the progenitor of this short GRB resides at large distance ($z\ge1$) or far away from its host galaxy, as for the case of a merging binary system with high kick velocity and/or long coalescing time. Another intriguing feature of this short GRB is that its 
X-ray afterglow showed an early (0.7 hours after the burst) steepening of the light curve. We explore possible origins of this steepening and discuss its implications. We present the observations and the performed data analysis in \S2,  the obtained results and the discussion in \S3 and \S4, respectively and a brief summary of the main results in \S5. All quoted errors are at the $90\%$ confidence level, unless specified.

\section{Observations and data analysis}

\subsection{Swift observations}

\subsubsection{BAT}

GRB 061201 was discovered with the Swift coded mask Burst Alert Telescope (BAT, Barthelemy et al. \cite{barthelmy2005}) on 2006 December 1  at $t_0$=15:58:36 UT (Marshall et al. \cite{marshall2006}). 
The BAT event data were analyzed following standard procedures (Krimm \cite{krimm2004}) with the BAT analysis software included in the HEASOFT distribution (v.6.1.2).

The burst had a duration of $T_{90}=0.9\pm0.1$ s in the 15--150\,keV band and two main peaks, one starting at the trigger time $t_0$ and the other peaking 0.8 s later (Fig.1). We find that the flux in the 15--25\,keV energy band lagged the 50--100\,keV flux by $2.7^{+3.3}_{-2.4}$ ms and the one in the 25--50\,keV energy band lagged the 100--350\,keV flux by $-0.8^{+1.5}_{-1.3}$ ms. These values are consistent 
with the typical (negligible) temporal lag values measured for short GRBs (Gehrels et al. \cite{gehrels2006}). We find no evidence of softer, extended emission at late times (e.g. Norris \& Bonnel \cite{norris2006}). 

The time integrated spectrum from $t_0$ s to $t_0+0.9$ s in the 15--150\,keV  energy range can be modelled by a simple power law as $F(E)\propto E^{-\alpha}$, with best fit 
energy spectral index $\alpha=-0.30\pm0.15$ (photon index $\Gamma=0.7$) and $\chi^2=19.8$ for 25 degrees of freedom. 
This value is consistent, within errors, with the results from the 20 keV -- 3 MeV Konus-Wind observations (Golenetskii et al. \cite{golenetskii2006}). The latters provide a time-integrated spectrum of GRB 061201 that was well fitted by a power law with exponential cutoff model, $F(E) \propto E^{-\alpha} e^{-(2-\alpha)E/E_p}$, with $\alpha = 0.36^{+0.40}_{-0.65}$ and $E_p = 873^{+458}_{-284}$ keV (Golenetskii et al. \cite {golenetskii2006}). 

The 15--150\,keV fluence observed by BAT was $(3.4\pm0.2)\times10^{-7}$ erg cm$^{-2}$ s$^{-1}$. The 20 keV-30 MeV fluence from Konus Wind data was $F_{20 keV - 3 MeV}=5.3\times10^{-6}$ erg cm$^{-2}$ (Golenetskii et al. \cite{golenetskii2006}).
During the burst there was clear evidence of a hard-to-soft evolution from 0.5 s after the trigger to the end of the burst (Fig.\ref{hr}).

   \begin{figure}
   \centering
   \includegraphics[angle=-90,width=9cm]{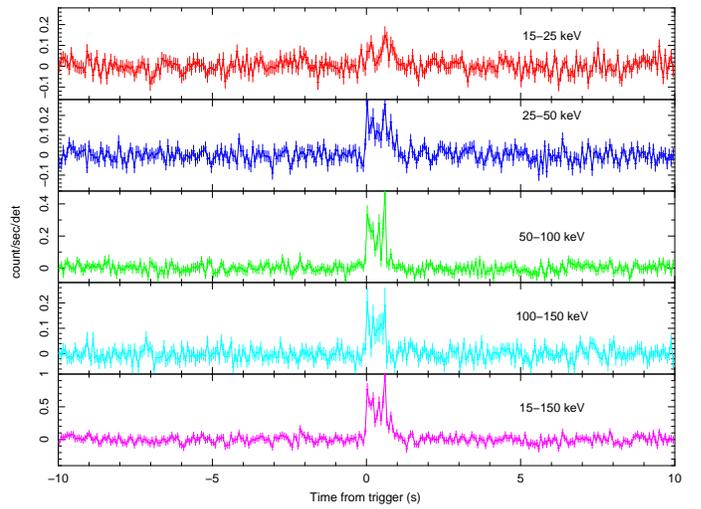}
      \caption{
BAT count rate (counts/sec/det) versus time since trigger for several energy bands. No extended emission is evident in the softer energy bands (e.g. 25--50\,keV) for this short GRB.
              }
         \label{lc}
   \end{figure}

   \begin{figure}
   \centering
   \includegraphics[angle=-90,width=9cm]{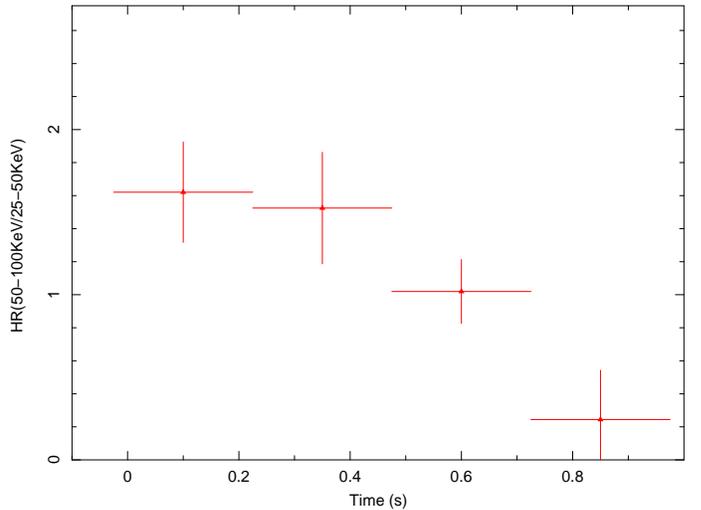}
      \caption{
Time evolution of the hardness ratio during the burst. A hard-to-soft trend is apparent starting from the second of the two peaks present in the burst light curve. 
              }
         \label{hr}
   \end{figure}

\subsubsection{XRT}

Swift slewed immediately to the burst and the X-Ray Telescope (XRT, Burrows et al. \cite{burrows2005}) began data acquisition at $t_0$+86 s in Windowed Timing mode. Between $t_0$+99 s and $t_0$+1.1~day the XRT was operated in Photon Counting mode. The XRT data were processed following standard procedures (Capalbi et al. \cite{capalbi2005}\footnote{http://swift.gsfc.nasa.gov/docs/swift/analysis/}), by using the last version of the dedicated XRT pipeline (xrtpipeline v 0.10.6). Grade filtering was applied by selecting the 0-2 and 0-12 ranges for the WT mode and PC mode data, respectively.  

Using 699 s of overlapping XRT Photon Counting mode and UVOT (UV-Optical Telescope,  Roming et al. \cite{roming2005}) V-band data,
we find an astrometrically corrected X-ray position (using the XRT-UVOT
alignment and matching to the USNO-B1 catalogue, see Goad et al. \cite{Goad2007})
which is R.A.= 22$^h$ 08$^m$ 32.23$^s$, Dec.=-74$^d$ 34$'$ 49.1$''$ (J2000.0) with an
uncertainty of $1.7''$ (radius, $90\%$ containment). This is 1.5$''$ from the initial XRT position  (Perri et al. 2006).


For the WT mode data, the afterglow signal was extracted from a rectangular region centered on the source position, with 40 pixels of width and 20 pixels of height. Due to the short exposure time in this mode (12 s), the number of photons was insufficient for a spectral analysis. 
For the PC mode, spectra and light curves were extracted from a circular region of 10 pixels of radius centered at the source position. From 99s to 719s after the burst, the PC mode data were above 0.5 cts/s, the threshold above which pileup effects become significant. In this time interval, we thus extracted the source photons from an annular region with 3 pixels and 20 pixels as inner and outer radii, respectively, in order to avoid pileup effects. 
Ancillary files were generated for each analyzed temporal region with the {\it xrtmkarf} task applying 
corrections for the PSF losses and CCD defects. The latest response matrices distribution (v8) was used. The count rate was corrected for effective area loss due to the presence of bad pixels in the extraction region. 

The extracted 0.3--10.0\,keV XRT light curve (Fig.\ref{lc}, Tab.\ref{xrtlc}) shows a fading behavior in the two covered temporal intervals after the burst, that is from 86 s to 719 s and from 1.5 hr to 34.3 hr.  The decay rate of the light curve in the two intervals featured a steepening. Assuming a broken power law model, where $F(t)\propto  t^{-\delta_1}$ for $t<t_b$ and $F(t)\propto t^{-\delta_2}$ for $t>t_b$, we found a good agreement with the data ($\chi^2=28.2$ with 28 degrees of freedom), with an early decay index $\delta_1=0.54\pm0.08$, temporal break at $t_b=2386^{+584}_{-521}$ s and late time decay index of $\delta_2=1.90\pm0.15$ (errors are at 1 $\sigma$ confidence level).

   \begin{figure}
   \centering
   \includegraphics[angle=0,width=9cm]{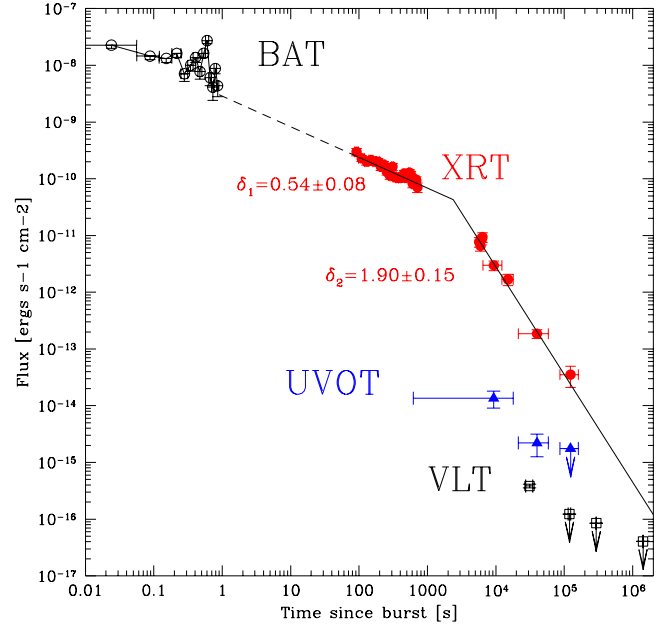}
      \caption{The 0.3--10.0\,keV unabsorbed flux light curve extracted from the BAT (open black circles, extrapolated to 0.3-10 keV range) and XRT (filled red circles) data sets as well as the optical flux from UVOT data ($UVW1$ filter, filled blue triangles) and from VLT data ($I$ band filter, open squares), not corrected for Galactic extinction ($E(B-V)=0.08$). The plotted model (solid line) is a broken power law fitted to the XRT data only. This gives a best temporal break of $t_b=(2.39^{+0.58}_{-0.52})$ ks. The dashed line is the extrapolation of the best fit model up to the temporal range covered by BAT.}
         \label{lc}
   \end{figure}

The 0.3--10.0\,keV energy spectrum was extracted in two epochs, before and after the temporal break. For both intervals we found a good agreement of the data with an absorbed power law model as 
$F(E)\propto E^{-\alpha_X}e^{-\sigma_{ph}(E)N_H}$, where $\sigma_{ph}(E)$ is the photoelectric cross section and $N_H$ is the equivalent hydrogen column density. 
Before the break, we found an energy spectral index of $\alpha_X=0.4\pm0.2$ and an equivalent hydrogen column density of $N_H=(10\pm6)\times10^{20}$ cm$^{-2}$ assuming that the absorbing matter is cold and with solar chemical abundance 
($\chi^2=15.4$ with 14 degrees of freedom). 
We note that the galactic hydrogen column in the direction of this burst is $N_{H,Gal}=4.75\times10^{20}$cm$^{-2}$ (Dickey and Lockman \cite{dl1990}), {\it i.e.} a factor of about two less than the total column required by the data. 
In the spectrum after the break we found an energy spectral index $\alpha_X=0.5\pm0.5$ ($0.5\pm0.2$ at 1 $\sigma$ confidence level) and column density upper limit of $N_H\le1.7\times10^{21}$ cm$^{-2}$ (90\% confidence level), consistent with the corresponding values before the break. 
In the absence of any evidence for spectral variations, we thus performed a simultaneous fit of the two spectra in order to better constrain the model parameters. 
In this fit all spectral parameters but the normalization were forced to be the same across different spectra. 
From the simultaneous fit we find $\alpha_X=0.4\pm0.2$ and a total column density of $N_H=(8\pm2)\times10^{20}$ cm$^{-2}$ ($\chi^2=22.8$ with 20 degrees of freedom). 

Finally we constructed the light curve from the trigger time (BAT data) to the latest time at which the X-ray afterglow was detected (XRT data). As the BAT and the XRT data cover different energy bands, we extrapolated the BAT data into the XRT energy  range (0.3--10.0\,keV). To this end, we fitted the BAT extracted spectrum by assuming a simple power law spectral model plus an absorbing column $N_H$ fixed at the value obtained from XRT data analysis.  We confidently extrapolated the BAT best fit model down to 0.3--10\,keV since we know that, from the Konus-Wind observations, the peak energy for this burst is at energies much higher than the BAT energy range.  We computed the unabsorbed 0.3--10\,keV flux and compared it with the corresponding observed (average) count rate, finding 1 c/s$_{BAT}=2.6\times10^{-8}$erg cm$^{-2}$ s$^{-1}$. 
The XRT count rate (corrected for effective area loss) was converted into flux by assuming the best fit spectral model obtained from the XRT data and comparing the unabsorbed 0.3--10.0\,keV flux with the relative observed (average) count rate per second. We found that 1 c/s$_{XRT}=6.8\times10^{-11}$ erg cm$^{-2}$ s$^{-1}$ (Fig.\ref{lc}). 



   \begin{figure}
   \centering
   \includegraphics[angle=-90,width=9cm]{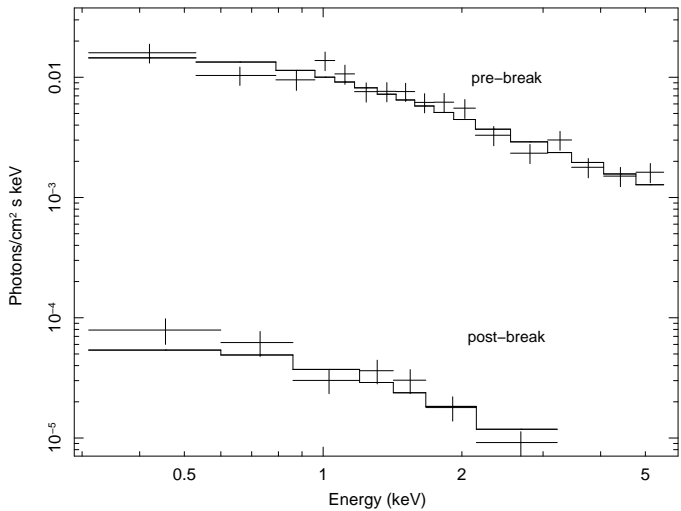}
      \caption{X-ray afterglow spectra extracted from XRT data taken in PC mode before and after the temporal break. No spectral variation is evident. The solid line is the best fit model from simultaneous fitting of the two spectra (see \S 2.1.2). 
              }
         \label{spec}
   \end{figure}

\begin{table}
\caption{The 0.3-10 keV unabsorbed flux light curve.}
\begin{tabular}{ccc}
\hline
 $T_{start}^a$ & $T_{end}^a$   & Flux \\
     (s)    & (s)       &  ($10^{-12}$erg cm$^{-2}$ s$^{-1}$)    \\
\hline
          86    &          98     &               $298\pm43$ \\
          99    &         119     &              $230\pm28$ \\
         119    &         139     &              $200\pm26$ \\
         139    &         159     &             $215\pm27$ \\
         159    &         179     &             $200\pm26$ \\
         179    &         199     &              $200\pm26$ \\
         199    &         219     &             $163\pm23$ \\
         219    &         239     &              $178\pm24$ \\
         239    &         269     &             $133\pm17$ \\
         269    &         299     &              $114\pm16$ \\
         299    &         319     &             $163\pm23$ \\
         319    &         349     &              $109\pm16$ \\
         349     &        379      &            $104\pm15$ \\
         379    &         409     &            $104\pm15$ \\
         409   &          439    &                 $109\pm16$ \\
         439    &         469     &             $123\pm17$ \\
         469    &         499     &            $104\pm15$ \\
         499   &          529    &                $118\pm16$ \\
         529   &          559    &               $128\pm17$ \\
         559    &         589     &                $118\pm16$ \\
         589   &          629    &                 $81.5\pm12$\\
         629    &         669     &           $100\pm13$ \\
         669   &          699    &                $94\pm14$ \\
         699   &          729    &                $71\pm13$ \\
        5579   &         5809    &               $7.7\pm1.5$\\
        5809   &         6079    &            $6.6\pm1.3$ \\
        6079   &         6289    &               $8.5\pm1.6$\\
        6289   &         6479    &               $9.4\pm1.8$\\
        6479   &        12230    &             $3.0\pm0.6$\\
	12230  &         17930   &             $1.7\pm0.4$\\
   21390       &         58440   &              $0.18\pm0.03$\\
   86430       &        160500   &                $0.03\pm0.01$\\
\hline
$^a$ Time from trigger
\end{tabular}
\label{xrtlc}
\end{table}

\subsubsection{UVOT}

The first UVOT finding chart was obtained starting at $t=t_0+$86 s. A faint  source was detected in the $U$, $UVW1$, $UVM2$, $UVW2$ and white filters. The highest signal to noise ratio detections were obtained with the $UVW1$ ($0.27 \mu$m) filter and with the white filter. Using the 290 s exposure in the white filter, the best fit position of GRB 061201 is R.A.=22$^h$ 08$^m$ 32.4$^s$ and Dec.=-74$^{\circ}$ 34$'$ 47.1$''$ (J2000) with $90\%$ confidence error of $1''$. This is 0.67$''$ and 0.22$''$ away in R.A. and Dec. respectively from the original UVOT position (Holland \& Marshall 2006).

The source count rate has been obtained from a circular region of aperture of 2.5$''$ and the background has been computed in an annulus region with inner radius of 15$''$  and outer radius of 25$''$. This aperture size was chosen to improve the signal to noise ratio for a weak source compared to the standard apertures for UVOT photometry. Aperture corrections were computed using stars in the image to make the rates correspond to the standard UVOT photometry apertures of $6''$ for the visible filters and $12''$ for the white and $UV$ filters.

In Table \ref{tab_log} we quote the magnitude obtained for each filter at several epochs with errors at 1 $\sigma$ and $90\%$ confidence level upper limits. The only filter for which we could confidently compute a timing analysis was $UVW1$. Thus, we fitted the $UVW1$ data at times later than the X-ray break with a simple power law $F(t)\propto t^{-\delta}$. The $90\%$ confidence power-law 
decay index is within the range $\delta=1.2\pm0.6$. This is slightly flatter than the X-ray data, but still consistent with them. No detection was found in either the $V$ or $B$ filter (see Table \ref{tab_log}). This could be due to the short total exposure with the $B$ filter (200 s) around $t_0+6000$s, and the lower sensitivity of the $V$ filter compared to that of the other filters, for bursts that do not suffer from extinction (see also \S 2.3).


\begin{figure*}
   \centering
  \includegraphics[width=15cm]{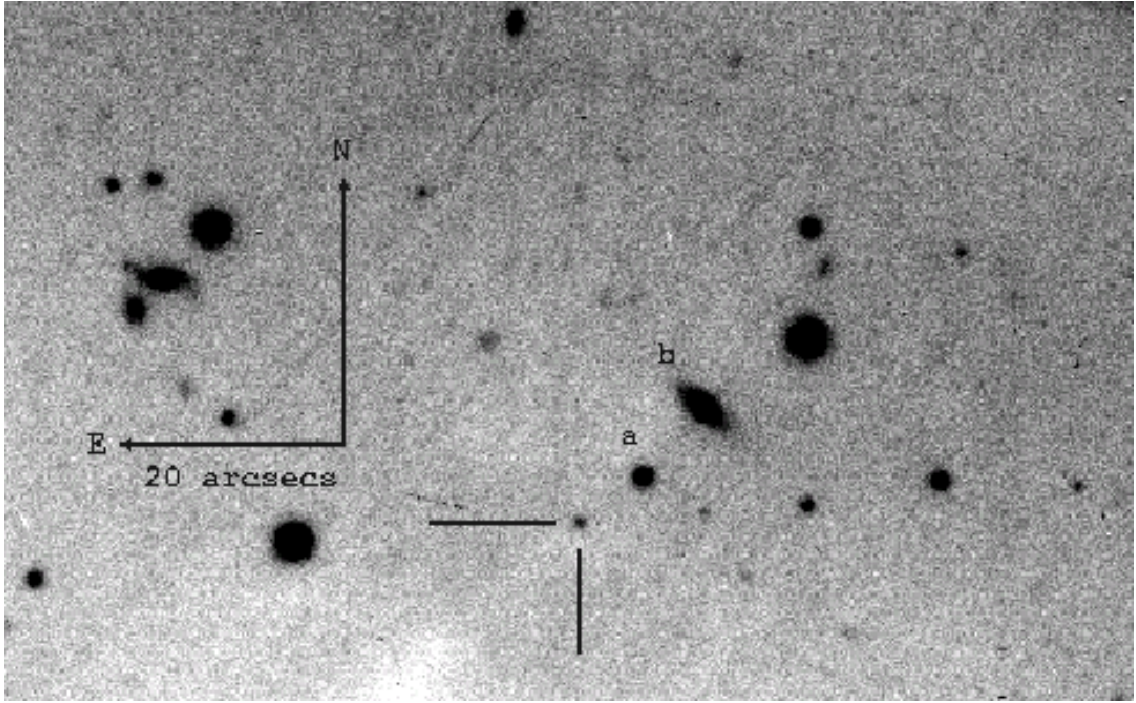}
  \caption{VLT finding chart for GRB\,061201. The afterglow is marked with solid lines. Objects ``a'' and ``b'' are the candidates host galaxies studied in the present paper. See section 2.2 for details.}
\label{finding_chart}
\end{figure*}

\subsection {VLT observations}

We observed GRB\,061201 with the ESO$-$VLT at six different epochs, starting about 8.6 hours after the burst. All the
observations were carried out by using the FORS1 and FORS2 cameras in both imaging and spectroscopic mode. All nights were photometric, with seeing in the 1.1$''$-1.7$''$ range. Image reduction was performed by following the standard procedures: subtraction of an averaged bias frame and division by a normalized flat frame. Observations in imaging mode were carried out in the $R$ and $I$ bands. The photometric calibration was achieved against Landolt standard stars, observed at different nights. Astrometric solutions were computed by using the USNO B1.0 catalogue\footnote{http://www.nofs.navy.mil/data/fchpix/}. Our spectra were taken with resolution of 8 \AA~ in the wavelength range from 4000 to 9000 \AA . We always used a 1$''$ slit and grating 330V. The extraction of the spectrum was performed within the IRAF\footnote{\textit{IRAF} is distributed by the National Optical Astronomy Observatories} environment. Wavelength and flux calibration of the spectra were obtained by using the helium-argon lamp and observing spectrophotometric stars. In Table \ref{tab_log} we present a complete log of our VLT observations. 

\subsubsection{Optical photometry}

$I$ and $R$ band observations of GRB\,061201 were obtained about 8.6 hours after the burst (D'Avanzo et al. \cite{davanzo2006a}). We revealed a faint candidate with $I = 22.36 \pm 0.08$ mag and $R = 23.05 \pm 0.12$ mag at R.A.=22$^h$08$^m$32.09$^s$ and Dec.= -74$^d34'47.08''$ (J2000) with uncertainty of 0.2$''$, consistent with that previously reported by Holland \& Marshall (\cite{holland2006}). We note that this is the most
accurate position available. In our second epoch of observation (33.1 hours after the burst) we could not detect the candidate down to a limiting magnitude of $I > 23.6$ ($3\sigma$ confidence level). This confirmed that the object was the optical afterglow of GRB\,061201 (D'Avanzo et al. 2006b). Subsequent $J$ band observations with the Southern Astrophysical Research telescope (SOAR), taken 10.2 and 33.6 hours after the burst, revealed a similar fading (Haislip et al. 2006). After correcting for the  Galactic extinction, that is $E(B-V)=0.08$ (Schlegel, Finkbeiner \& Davis \cite{schlegel1998}), we computed the afterglow color for GRB 061201 of $R-I=0.63\pm0.14$. 

We continued to monitor the field of GRB\,061201 until 2006 Dec 18 (about 16.4 days after the burst): no underlying galaxy was detected down to limiting magnitudes $R > 24.9$ and $I > 24.8$ ($3\sigma$ confidence level, see Tab. \ref{tab_log}). Assuming a power law slope ($F(t) = F_0t^{-\delta}$) we constrained the flux decay index to be $\delta\ge0.85$ from the first $I$ band flux upper limit . 

On 2007 May 22 we performed another exposure of the field with the $R$ filter. No host galaxy was detected down to $R>25.9$ mag with 1.6 hours of exposure (Tab. \ref{tab_log}).

\begin{table*}
\caption{Observation log and photometry, not corrected for Galactic extinction. }
\centering
\begin{tabular}{ccccccc}
\hline
UT observation    &  $t-t_0$ & Exposure                             &   Instrument &  mag                & Filter\\
        &      (days)       &(s)                               &              &                     &       \\
\hline
2006 Dec 1.70269 &0.037   &290  &  Swift/UVOT &  $20.90\pm0.30$ & White \\
2006 Dec 1.70369 &0.038   &214  &  Swift/UVOT & $20.86\pm0.54$  &  U  \\
2006 Dec 1.70569 &0.040   &202  &  Swift/UVOT & $>21.14$	   &   B  \\
  2006 Dec 1.70769  &0.042   &122  & Swift/UVOT & $19.94\pm0.41$  & UVW2  \\
2006 Dec 1.73669 &0.071   &1111 &  Swift/UVOT & $>21.23$           &     V \\
2006 Dec 1.77269 &0.107   &519  &  Swift/UVOT & $20.75\pm0.36$  & UVW1  \\
2006 Dec 1.78869 &0.123   &676  &  Swift/UVOT & $21.44\pm0.54$   & UVM2  \\
2006 Dec 2.02461  & 0.359 &$14 \times 260$  & VLT/FORS2    & $22.36 \pm 0.08$    & $I$   \\
2006 Dec 2.03886  & 0.373 &$3  \times 260$  & VLT/FORS2    & $23.05 \pm 0.12$    & $R$   \\
2006 Dec 2.12769 &0.462   &11529&  Swift/UVOT & $22.72\pm0.47$   & UVW1\\  
2006 Dec 3.09469 &1.429   &19928&  Swift/UVOT & $>22.97$	     & UVW1 \\
2006 Dec 3.04453  & 1.379 &$20 \times 260$ & VLT/FORS2    & $> 23.6 (3\sigma)$  & $I$   \\
2006 Dec 5.05761  & 3.392  &$40 \times 3 \times 20$  & VLT/FORS2    & $> 24.0 (3\sigma)$  & $I$   \\
2006 Dec 15.03991 & 13.374 &$2  \times 180 + 1 \times 60$  & VLT/FORS2    & $> 24.9 (3\sigma)$  & $R$   \\
2006 Dec 18.04937 & 16.383 &$12 \times 120$   & VLT/FORS2    & $> 24.8 (3\sigma)$  & $I$   \\
2007 May 22.29685 & 171.631 & $24\times240$    & VLT/FORS2    & $>25.9 (3\sigma)$  & $R$   \\
\hline
2006 Dec 2.07588  & 0.410 &$2 \times 1800$  & VLT/FORS2    & $-$    & $300V+GG375$   \\
2006 Dec 15.06615 & 13.400& $2 \times 1200$  & VLT/FORS1    & $-$    & $300V+GG375$   \\
\hline
\end{tabular}
\label{tab_log}
\end{table*}

\subsubsection{Optical spectroscopy and host galaxy candidates for GRB061201}

After the identification of the optical counterpart we took an optical spectrum of the afterglow in two epochs (see Table 1). Unfortunately, due to visibility constraints, we could integrate only for one hour, resulting in a low signal-to-noise spectrum.  This was characterized by a very weak continuum with no distinguishable emission nor absorption features (except for telluric ones). Therefore, no redshift could be determined.

As seen from Figure \ref{finding_chart}, two bright objects (object ``a'' and ``b'') are present close to the afterglow position. 
We took an optical spectrum of both, in order to investigate their nature and look for possible relations with GRB\,061201.

Object ``a'' is located 7.5$''$ NW of the position of the optical afterglow and has a stellar-like profile. Our photometry shows that this object is particularly red, with $(R-I)=1.00 \pm 0.04$ mag, suggesting that it could be a main sequence late$-$type star. This hypothesis was confirmed through the analysis of the spectrum of this object, characterized by a strong Na line and several TiO bands, which are typical features of a M0$-$M2 main sequence star.

\begin{figure}
   \centering
  \includegraphics[angle=-90,width=9cm]{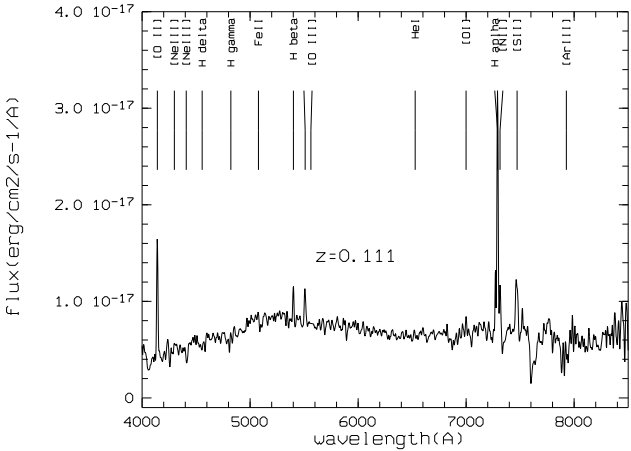}
  \caption{VLT spectrum of the nearest galaxy (object ``b'' in Fig.\ref{finding_chart}) at z=0.111.}
\label{VLT_spec}
\end{figure}

Object ``b'' is located 17$''$ to the NW of the afterglow and is clearly extended. Our VLT spectrum (Fig. \ref{VLT_spec}) shows several emission lines among which we identified the [O\,II] and H$\alpha$ lines redshifted at z=0.111. This result is consistent with previous findings (Berger et al. \cite{berger2006a}).  
As indicated by the prominent emission lines, star formation is still present in this galaxy. For the [O\,II] and H$\alpha$ emission lines we measured luminosities of $9.92 \times 10^{39}$ and $1.83\times 10^{40}$ erg s$^{-1}$ (corrected for slit loss). This corresponds to an unobscured star formation rate of 0.14 $M_{\odot}$ yr$^{-1}$ (Kennicutt 1998), which corresponds to about 1.2 $M_{\odot}$ yr$^{-1}L_*^{-1}$ once normalized to $L_*$ (assuming $M_B = -18.6$ for the galaxy \footnote{as determined in the USNOB1 catalog and corrected from Galactic extinction} and ${M_B}^* = -21$  for field galaxies). This is significantly less than the typical SFR observed in long GRB host galaxies (Christensen et al. \cite{christensen2004}). 
Comparing with other short GRBs host galaxy SFRs, this is lower than that of the host of GRB 051221A by a factor of 3.3 (Soderberg et al. 2006) but much larger than that of the hosts of the short GRBs GRB\,050509B (Bloom et al. \cite{bloom2005}, Gehrels et al. \cite{gehrels2006}) and GRB 050724 (Berger et al. \cite{berger2005}), by factors of more than about 20 and 60 respectively. We also note that our SFR value is of the same order as that measured for GRB\,050709, a short GRB for which the association to a star-forming galaxy is secure (Covino et al. \cite{covino2006}).  

We finally tested the hypothesis that the host galaxy is coincident with the optical afterglow but too faint to be detected. If the host galaxy of GRB 061201 were similar to that of GRB 050709 it would have an absolute magnitude of 
$M_R=-18.3$. In this case, in order to not be detected in the $R$ band down to 25.9 mag, the host galaxy should be at redshift higher than $\sim1$. If GRB 061201 host were a normal galaxy, its $R$ band upper limit would indicate a redshift above $\sim1.5$ (e.g. Berger et al. 2007).


\subsection{Multiwavelength analysis}

Despite the fact that the afterglow was detected up to the $UVW2$ filter, no $B$ and $V$ bands emission was found.  In addition, the afterglow $R-I$ color is quite red, measuring $0.63\pm0.14$ mag (after correcting for the Galactic extinction). We further investigate these observational aspects by extracting an optical-to-X-ray spectral energy distribution (SED). 

Due to the temporally sparse sample of optical data and the large uncertainty on the optical decay rate, we identified two epochs at which the optical afterglow was detected at least with three different filters nearly simultaneously, thus minimizing error propagation during the extrapolation at a common epoch for the SED extraction.  These epochs are $t=t_0+0.040$ day (3.4 ks) for the $V$, $B$, $U$, $UVW1$, $UVW2$ filters and $t=t_0+0.463$ day (40.0 ks) for the $I$, $R$ and $UVW1$ filters (Tab. 1). After correcting for the Galactic extinction ($E(B-V)=0.08$), we rescaled the corrected magnitudes and the unabsorbed X-ray flux to the selected epochs. We then converted the magnitudes into fluxes and we construct the optical-to-X-ray SED for each epoch. 

During the first epoch (3.4 ks), that is just after the observed temporal break, we find that the optical-to-X-ray SED is well described by a simple power law with spectral index $\alpha_{opt,X,1}=0.5\pm0.1$. The latter is consistent with the spectral index we find from the X-ray analysis based on the averaged spectrum extracted after the temporal break (see \S 2.1.2). The $V$ and $B$ bands upper limits are consistent with not being deep enough for the detection (Fig. 7). During the second epoch (40 ks), we find that to connect the X-ray flux to the optical fluxes, a steeper power law is required, with $\alpha_{opt,X,2}=0.82\pm0.15$. The $I-R$ color is consistent with this spectral slope. However, the large uncertainty on the $UVW1$ flux prevented us from firmly excluding any possible reddening (Fig.\ref{SED}).


\section{Results}

\subsection{The origin of the break in the X-ray light curve}

The spectral and temporal properties of the prompt emission of GRB 061201 that we find, are consistent with the identification of this burst as a short GRB. The most interesting feature for this GRB is perhaps the clear X-ray light curve steepening (temporal break) observed $\sim0.7$ hours after the burst. This is the best example of a temporal break in the X-ray afterglow of a short burst to date. Only one  other short burst so far showed convincing evidence of a temporal break in the X-ray light curve (GRB 051221A, Burrows et al. \cite{burrows2006}, Soderberg et al. \cite{soderberg2006}). 

A simultaneous light curve steepening at all frequencies and with similar post-steepening decay rate is generally interpreted as evidence of a jet-collimated afterglow (e.g. Sari et al. 1999). For GRB 061201, the achromatic nature of the temporal break could not be firmly established.
The UV light curve shows a decay after the break that is somewhat shallower than the X-ray decay but consistent within the uncertainties, while the $I$ band observations provide only an upper limit, yet consistent with the X-ray decay rate after the break. 

We checked whether the observed steepening of the light curve may be due to a synchrotron characteristic frequency crossing the energy range of the XRT. If this were the case, a change in the X-ray spectral index of $\sim0.5$ simultaneous with a steepening of the light curve would be expected (Sari et al. \cite{sari1998}). 
The large uncertainty on the post-break spectral index prevented us from reaching firm conclusions. Nevertheless,  the best fit post-break spectral index value is consistent with the spectral index before the break and the associated uncertainty excludes any spectral variation at $68\%$ confidence level. A more contrainings result comes from the observed XRT light curve steepening, that corresponds to an increase of the decay index of $\Delta\delta=1.4\pm0.2$ that is much larger than expected in the case of a synchrotron frequency crossing the X-ray energy range (e.g. Panaitescu 2006), definitively excluding this scenario. 

The temporal and spectral properties of the X-ray afterglow are consistent with the relationships predicted by the fireball model in the context of a collimated emission.
In fact, both the X-ray spectral index ($\alpha_X=0.4\pm0.1$ at 1 $\sigma$ confidence level) and the temporal decay before the break ($\delta_1=0.54\pm0.08$) are consistent with an expanding fireball in a slow cooling regime, with the cooling frequency $\nu_c$ still above the X-ray energy domain (Sari et al.\cite{sari1998}). Thus, the expected energy spectral index is $\alpha_X=(p-1)/2$ from which we can derive the electron spectral index $p=1.8\pm0.2$. 
The predicted decay rate after a temporal break due to the presence of a jet is expected to be identical to $p$ (Sari et al. \cite{sari1999}): this is in agreement with the observations. 

\begin{figure}
   \centering
  \includegraphics[angle=0,width=9cm]{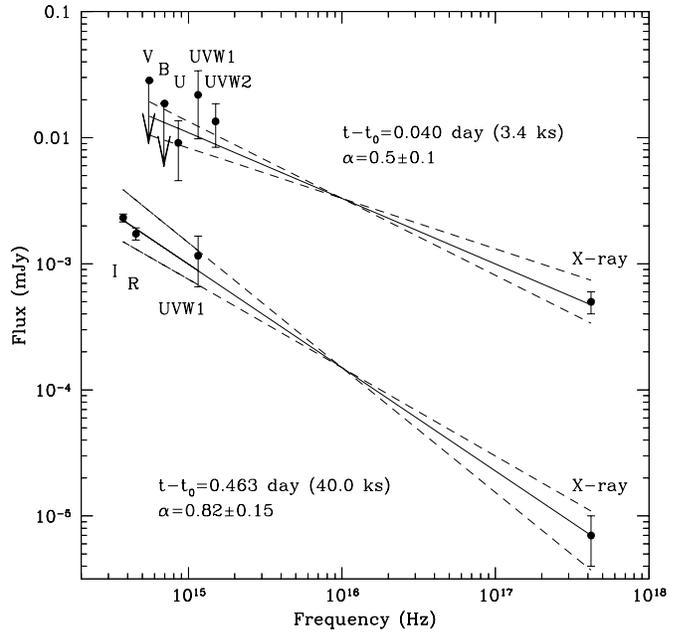}
  \caption{Observed spectral energy distributions at two different epochs (both after the temporal break).  The optical fluxes have been corrected for the Galactic extinction. The solid lines are the best fit power law models find at each epoch, where $\alpha$ is the best fit spectral index. The dashed lines define the expected range of values at $90\%$ confidence level.}
\label{SED}
\end{figure}

The spectral index obtained from the optical-to-X-ray spectral analysis 3.4 ks after the burst, that is just after the break, is consistent with the one obtained from both the pre and post break averaged X-ray spectra. However, it shows a possible steepening of the spectrum at late times (10 hours after the break, Fig.\ref{SED}).  This is difficult to explain in a jet scenario as the crossing of the cooling frequency towards lower energies, since $\nu_c$ is not expected to vary after the jet break (Sari et al. 1999).


We further investigate other possible scenarios. In particular, the pre-break X-ray light curve can be identified with the shallow decay (with typical decay index $\delta\le0.5$) observed in several long GRB X-ray afterglows in the Swift era, followed by the 'normal' decay after the break (with typical decay index $\delta\sim 1$, see Liang et al. 2007, Willingale et al. \cite{willingale2007}, O'Brien et al. \cite{obrien2006}). As observed so far for long GRBs, no spectral variation is present across the break: this is also the case of GRB 061201. The origin of the 'shallow phase' is still a mystery (e.g. Zhang 2007). On the contrary, the 'normal phase' is expected to follow the standard afterglow closure relationships. In this scenario, the observed spectral steepening of the optical-to-X-ray SED observed in two epochs after the temporal break, could be explained by the crossing of the cooling frequency. We note, however, that the X-ray decay index of GRB 061201 during the 'normal' phase (post-break) is much steeper than typical values observed so far for long GRBs (Liang et al. 2007), and its consistency with the electron spectral index measured from the X-ray spectral analysis should be considered as a coincidence.



\subsection{Distance estimates}

For this GRB no underlying host galaxy was detected after the fading of the optical afterglow and no firm redshift could be obtained. 
The Swift/UVOT detection in the $UVW2$ filter proves that the source is at $z\le1.7$ (we conservatively considered the red limit of the filter). The optical afterglow position lies $8.5'$ from the center of the rich ACO S 995 cluster of $22'$ diameter (Abell et al. \cite{abell1989}) for which no redshift was found in the literature{\footnote{We note that the redshift z=0.237 estimated for the ACO S 995 cluster by Blondin et al. (\cite{blondin2006}) has been confirmed to be wrong (Blondin, private communication).} . The spectra of nine apparent members of this cluster were obtained by Berger et al. (\cite{berger2006a}) within a redshift range of $0.0843<z<0.0877$, concluding that the redshift of ACO S 995 is z=0.0865. 

Our VLT images revealed the stellar nature of the nearest object to GRB 061201 (``a'' in Fig.\ref{finding_chart}) in sky projection but, more importantly, place tight constraints on the presence of an underlying host galaxy. We demonstrated that, if the host galaxy of GRB 061201 were similar to the host of the short GRB 050709, it should be at $z\ge$1 in order to not be detected. Moreover, if we compare our magnitude limit with a sample of faint galaxies at known redshifts (Benitez et al. 2004, Berger et al. 2007), GRB 061201 would be likely at redshift above 1.5. We find that the nearest galaxy (with offset of 17$''$) is at z=0.111 (object ``b'' in Fig.\ref{finding_chart}), confirming previous results from Berger et al. (2006b). We also demonstrated that this galaxy shows evidence of star formation activity. 

We tested the consistency of the association of GRB 061201 with the ACO S 995 cluster and with the z=0.111 nearby galaxy.
From a collection of orbital parameters of various eccentric binary systems by Champion et al. (\cite{champion2004}), it appears that these systems can be split into those that will and will not coalesce within a Hubble time.  The former group into which GRB 061201 should be put by definition, is characterized by coalescing times $\tau$  within 0.1 and 3 Gyr. If the nearby galaxy (object ``b'' in Figure 6) at z=0.111 is the host of GRB 061201, its angular offset would imply a projected distance of 42 kpc and thus a kick velocity of $\ge10-400$ km s$^{-1}$. If GRB 061201 was instead at the redshift of the ACO S 995 galaxy cluster at z=0.0865, it would be at $\ge0.9$ Mpc from its center, requiring a high kick velocity $\ge300$ km s$^{-1}$ up to uncomfortably high lower limits (few $\times 10^3$ km s$^{-1}$) if $\tau\sim$0.1 Gyr. Therefore, for a coalescing time $\tau\ge1$ Gyr, both the ACO S 995 galaxy cluster and the z=0.111 galaxy are consistent with hosting GRB 061201, while for $\tau\le1$ Gyr, only the galaxy at z=0.111 should be considered as the host candidate.  

\begin{table*}
\caption{Energy estimations for different redshift.}
\centering
\begin{tabular}{ccccc}
\hline
Redshift    &  $E_{\rm iso}$ & n &$\theta_{\rm jet}$ & $E_{\gamma}$  \\
	    &  ($10^{50}$erg)      & (cm$^{-3}$) &  (deg)    & ($10^{48}$erg)  \\
\hline
0.0865      &	0.8 & $10^{-3} \div 4\times10^{-2}$ & $1.3 \div 2.2$ & $2\times10^{-2} \div 6\times10^{-2}$ \\
\hline
0.111	    &1.4   & $10^{-3} \div 3\times10^{-2}$ &  $1.2 \div 1.9$ & $3\times10^{-2} \div 8\times10^{-2}$ \\
\hline
$\sim1$	  &	140	& $8\times10^{-5} \div 2\times10^{-3}$  & $0.4 \div 0.6$  & $0.3 \div 0.8$\\
\hline
\end{tabular}
\label{ene}
\end{table*}

\subsection{Energetics}

Given the distance uncertainties, we computed the energetics of GRB 061201 at each proposed redshift, i.e. $z=0.0865$ and z=0.111. We also discussed the possibility of a positionally coincident host galaxy at z$\sim1$ (see \S 3.2). In the following calculations we assumed a flat Friedman-Robertson-Walker cosmology with Hubble constant of $H_0=71$ km s$^{-1}$ Mpc$^{-1}$, $\Omega_M=0.27$ and $\Omega_{\Lambda}=0.73$. 
In order to compute the total isotropic equivalent energy $E_{\rm iso}$ we used the 20 keV - 3 MeV Konus-Wind observed fluence (rather than the 15--150\,keV BAT fluence) which provides a fair approximation of the bolometric value (see \S 2.1.1). Results are quoted in Table \ref{ene}.  

We then make the hypothesis of a collimated emission, as the temporal and spectral properties of the X-ray observations suggest.  
From the rest frame time of the light curve break, we computed the allowed values of jet opening angle. This is possible since at the time of the break, the increasing relativistic beaming angle of the fireball emission $1/\Gamma(t)$ (where $\Gamma(t)$ is the Lorentz factor of the fireball) is expected to be equal to the geometrical opening of the jet. From this it is possible to derive the expression of the jet opening angle $\theta_{\rm jet}$ as a function of the parameters which $\Gamma(t)$ depends on: these are the relativistic outflow energy $E$, the density of matter $n$ in the GRB surroundings and the fraction of kinetic energy of the outflow that goes into the prompt emission radiated energy (efficiency) $\eta=E_{\rm rad}/E$ (Sari et al. \cite{sari1999}). We assumed that $E_{\rm iso}$ is an estimate of the prompt emission radiated energy and that $\eta=0.1$.  

We attempted to estimate the density of matter from the start time of the afterglow, that is, from the time $t_d$ at which the expanding fireball has swept up enough matter to start its deceleration: this is $t_d=0.1\times15(E/10^{50} $erg$)^{1/3}(1 $cm$^{-3}/n)^{1/3}(300/\Gamma)^{8/3}$ seconds (e.g. Vietri 2000). Assuming that the afterglow emission starts before the XRT observations starting time, that is $t_d\le 86/(1+z)$ s, we could infer a lower limit to the density from the inequality $n\ge 10^{-3}(15$s$(1+z)/t_d)^3(E/10^{50} $erg$)(300/\Gamma)^8$ cm$^{-3}$.  We note, however, that this estimation strongly depends on the cube of the unknown deceleration time $t_d$, that can be up to one order of magnitude shorter than the XRT observations starting time and, more importantely, on the eighth power of the uncertain value of the fireball Lorentz factor $\Gamma$ that for long GRBs is thought to be $\sim300$ (e.g. Molinari et al. 2007) and for short GRBs it might reach values down to 40 (Nakar 2007): we assumed here an average value of 170.

Another estimate of density lower limit can be achieved from the peak flux of the X-ray afterglow $f_{\rm max}=2.6$mJy$(1+z)(\epsilon_B/0.01)(E/10^{50}$erg$)(n/1$cm$^{-3})^{1/2}(D_{\rm L}/10^{27}$ cm$)^{-2}$, assuming that $f_{\rm max}$ is above the first XRT flux measure (that is the X-ray flux at $t=t_0+86$ s), where $\epsilon_{\rm B}$ is the fraction of energy that goes into magnetic field and $D_L$ is the luminosity distance (Sari \& Esin \cite{sari2000}).  These estimations however provide very low, thus useless, lower limits. 

We note that the ratio of the X-ray flux at time $t$ multiplied by $t$ to the 15--150\,keV prompt observed fluence $f_{\rm X,\gamma}$ depends on the density if $\nu_X<\nu_c$ as $f_{\rm X \gamma}=0.01 (1/\eta)(\epsilon_e/0.1)^{3/2}(\epsilon_B/0.01)(E/10^{50} $erg$)^{1/3}(n/1 $cm$^{-3})^{1/2}$ where $\epsilon_e$ is the fraction of total energy that goes into the electron population (Nakar \cite{nakar2007}). However, 
if we interpret the temporal break as a jet signature, the above expression for $f_{\rm X,\gamma}$ could not be applied since it assumes an isotropic energy release. 


Density upper limits can be measured through the synchrotron cooling frequency density dependence (Sari et al. 1998).  
Since the X-ray spectrum is consistent with $\nu_X<\nu_c$, by imposing this condition to the $\nu_c$ expression given by Sari et al. (1998) for the case of an adiabatic, isotropic expansion, we could derive a density upper limit as  $n<1.3\times10^{-3}(\epsilon_B/0.01)^{-3/2}(E/10^{50} $erg$)^{-1/2}(\nu_X(1+z)/1.2\times10^{18} $Hz$)^{-1}(t_d/0.028(1+z)$day$)^{-1/2}$ cm$^{-3}$. We fixed the time dependence at $t_d=t_{\rm jet}$, that is at the maximum time where we could assume a spherical geometry, and we assumed $\nu_X$ as corresponding to the energy of 5 keV.

The estimation of $\theta_{\rm jet}$ was obtained at each redshift assuming for the density of the environment the most constraining range of values find above ( that is, lower limits from the afterglow deceleration time and upper limits from the cooling frequency density dependence). We then computed the allowed values for the beaming corrected energy by applying the beaming correction factor $(1-cos\theta_{\rm jet})$ to the estimated range of $E_{\rm iso}$. Results are given in Table \ref{ene}.

\section{Discussion}

We have presented a multiwavelength study of the short burst GRB 061201. The most interesting feature of this burst is the presence of a clear steepening of the afterglow light curve observed in X-rays at $\sim0.7$ hour after the burst. This is probably the best evidence so far for a temporal break in a short GRB afterglow. We demonstrated that the temporal and spectral properties of the X-ray afterglow are consistent with a jet origin of this break, although optical data can not definitively confirm this scenario and other explanations are possible. 

The observation of jet breaks in short GRBs is of utmost importance because it opens the possibility to infer the degree of collimation of the jet and burst energetics which are ultimately related to the progenitor model  so we deeply investigated this possibility. While for long GRBs the stellar envelopes of the progenitor are expected to drive the release of energy into fairly highly collimated jets, numerical simulations of coalescing binary systems, that are thought to be among the progenitors of short bursts, predict a lower degree of collimation.
Short burst jet opening angles estimated so far were obtained for GRB 051221A, angle of  $\sim4-8$ degrees (Burrows et al. \cite{burrows2006}, Soderberg et al. \cite{soderberg2006}) based on multiwavelength observations (from radio to X-rays) and 
for GRB 050709, based however on few data points that provided a poorly constrained jet angle of about $\sim15$ degrees (Fox et al. 2005, but see also Watson et al. 2006). For GRB 050724, a jet angle of $\sim 8-12$ degree was claimed from radio and NIR observations (Berger et al. \cite{berger2005}). This measure is however still under debate since no break was observed in the X-ray light curve up to three weeks after the burst, implying a jet opening angle lower limit of 25 degrees (Grupe et al. \cite{grupe2006}, Malesani et al. \cite{malesani2007}). For comparison, long GRBs show evidence of jet opening angles that goes from 4 degrees (e.g. Frail et al. 2001) to 10 degrees (e.g. Guetta et al. 2005).
The uncertainty on the distance of GRB 061201 and on the density of the surrounding environment, prevented us from firmly estimating a jet opening angle. 
However, exploring different redshift and density ranges, we find that if the jet break interpretation is correct, GRB 061201 seems to be more collimated than those short GRBs for which evidence of collimation was detected so far, extending the jet opening angle for short GRBs below two degrees, and more collimated than many long bursts despite average predictions of binary merger models.

Alternatively to the jet interpretation, the observed break in the X-ray light curve can be identified with the transition from an early shallow decay to a 'normal' afterglow decay, as observed in several long GRBs in the Swift era. Several mechansims have been invoked to explain this phenomenology that go from refreshed external shocks due to continuous energy injection from a long-term central engine or from an ejecta with a wide Lorentz factor distribution, to X-rays echoes from dust scattering (see Zhang 2007 for a review). Despite the uncertainties on the interpretation, should the observed X-ray temporal break for GRB 061201 be identified with this phenomenology, all these models should be refined in order to include the class of short GRB progenitors that is expected to be different from that one of long GRBs. 

We checked whether GRB 061201 satisfies the $E_{\rm p,i}-E_{\rm iso}$ and $E_{\rm p,i}-E_{\gamma}$ correlations found for long GRBs (Amati et al. 2002, Ghirlanda et al. 2004), where $E_{\rm p,i}$ is the intrinsic peak energy of the $\nu F_{\nu}$ spectrum of the prompt emission. We find that, by rescaling the measured $E_{\rm p}=873^{+458}_{-284}$ keV from Konus-Wind observations (Golenetskii et al. \cite{golenetskii2006}) at each proposed redshift for this burst, the isotropic and beaming corrected energy are several orders of magnitudes lower than the values predicted by the correlations. This behavior is in line with past results for short GRBs (e.g. Amati \cite{amati2007}) and supports the idea of a different origin of long and short GRBs.

So far, short GRBs with measured spectroscopic redshift had isotropic equivalent energies $E_{\rm iso}$ ranging  between $10^{48}$ erg and $10^{51}$ erg (e.g. Fox et al. 2005). Five short GRBs with no clear host galaxy  association, but for which galaxies were detected within the XRT error circle, were claimed to reside at redshift above 0.7 and for these sources $E_{\rm iso}$ ranges between $10^{50}-10^{52}$ erg (Berger et al. \cite{berger2007}). 
In one case (GRB 051221A) the collimation factor was measured from solid evidence of a jet break in the X-ray afterglow light curve and the beaming corrected energy was $(1-5)\times10^{49}$ erg (Burrows et al. \cite{burrows2006}). If collimated, the short GRB 050709 would have a beaming corrected energy of $2\times10^{48}$ erg and GRB 050724 of $4\times10^{48}$ erg. Thus, so far, no short GRB has shown a released energy below $\sim10^{48}$ erg.   
For GRB 061201, the isotropic equivalent energies given in Table \ref{ene}, estimated assuming different redshifts, are within the observed range of values for short GRBs. However, if the jet break interpretation is correct, the beaming corrected energy estimated in the low redshift cases ($0.0865<z<0.111$) is about two order of magnitude less than the minimum energy estimation obtained so far for short GRBs. 
If instead GRB 061201 were residing at high redshift within a faint host galaxy with $R\ge25.9$ mag at $z\sim1$, its energy content, even if beaming corrected, would be more similar to previous short GRBs with estimated redshift. 

The lack of any detection of an underlying host galaxy after the fading of the optical afterglow may suggest some hints on the origin of this short GRB.  
One possibility is that GRB 061201 is far away from its host galaxy, therefore in a low density environment as can be the halo of the nearby galaxy at z=0.111 or the intracluster medium of the ACO S 995 cluster at redshift of z=0.0865 (see \S 3.2). In this case,  GRB 061201 may represents the first evidence of a (low energetic, if beamed) class of coalescing binary system, with high kick velocity and/or long coalescing time.  Another possibility is that the host galaxy is positionally consistent with the GRB but too faint to be detected, that
is, likely to be at redshift $z\ge1$ (see \S 2.2.2). Indeed, a fraction of 30\%-60\% of short GRBs with very faint host galaxy is thought to reside at $z\ge0.7$ (Berger et al. \cite{berger2007}), possibly supporting this scenario.

\section{Summary}

We carefully analyzed the X-ray and optical emission from the short GRB 061201. 
The most relevant results we found are: 

\begin{itemize}

\item
a clear steepening of the X-ray light curve at $\sim0.7$ hours after the burst trigger with post-break X-ray decay index $\delta\sim2$. 

\item
no evidence for a host galaxy after optical afterglow faded, down to $R\sim26$ mag; the closest galaxy has an offset of 17$\arcsec$ and is a star forming galaxy at z=0.111.

\end{itemize}

The distance of this burst is unknown. A possible association of this GRB whith the ACO S 995 cluster of galaxies at $z=0.0865$ has been suggested. Alternatively, it can be associated with the nearby galaxy at $z=0.111$. We discussed all these cases as well as the case of a positionally coincident host galaxy at $z\sim1$, too faint to be detected. 
If the X-ray light curve break were due to a jetted fireball, the range of opening angle values obtained assuming several scenarios (distance scales) would be below 2 degrees.

\begin{acknowledgements}

We thanks Daniele Malesani for useful discussions and the anonymous referee for his/her precious suggestions. This work is supported in Italy from ASI Science Data Center and by ASI grant I/024/05/0 and MIUR grant 2005025417.

\end{acknowledgements}

\end{document}